\documentclass[aps,floatfix,pra,superscriptaddress,twocolumn]{revtex4-1}
\usepackage{graphicx}
\usepackage[english]{babel}
\usepackage{amsmath}
\usepackage{amssymb}
\usepackage{tensor}

\newcommand{\vd}{\mathbf{d}}

\newcommand{\vbr}{\mathbf{r}}
\newcommand{\be}{\begin{eqnarray}}
\newcommand{\ee}{\end{eqnarray}}
\newcommand{\p}{\partial}

\def\ep#1{\langle #1 \rangle}

\begin{document}

\title{Vortices in Two-Dimensional Chiral Superfluids}

\author{Yan He}
\affiliation{College of Physics, Sichuan University, Chengdu, Sichuan 610064, China}

\author{Wenxing Nie}
\email{wxnie@scu.edu.cn}
\affiliation{College of Physics, Sichuan University, Chengdu, Sichuan 610064, China}

\begin{abstract}

We study the orbital angular momentum (OAM) $L_z$ of two-dimensional chiral $(p_x+ip_y)^{\nu}$-wave superfluids (SFs) in the presence of an axisymmetric multiply quantized vortex (MQV) with vorticity $k$ on a disk at zero temperature, in the framework of Bogoliubov-de Gennes (BdG) theory. Focusing on spectral asymmetry (or spectral flow), we find that $L_z=(k+\nu)N/2$ for any integer $\nu$ and $k$ in the Bose-Einstein Condensation (BEC) regime, where $N$ is the total number of fermions. While  in the weak-pairing Bardeen-Cooper-Schrieffer (BCS) regime, only for chiral $p+ip$-wave SF with $k=\pm 1$, $L_z=(k+\nu)N/2$ still holds. For chiral SFs with $\nu\ge2$ or $|k|\ge2$ in the BCS regime, the OAM $L_z$ is remarkably reduced from its ``full" value in the BEC regime.
However, the deviations differ in these two cases. For chiral SFs with $\nu\ge2$, $L_z$ is sharply suppressed in this ideal setting with a specular wall, while the suppression caused by the $|k| \ge 2$ vortex is moderate, which is core-size dependent.
Furthermore, for $p+ip$-wave SF with $k=-1$, the total OAM $L_z$ is zero, but the distribution $L_z(r)$ is nontrivial compared with that of vortex-free $s$-wave SF, in which the total OAM is zero as well.
For chiral SFs with $\nu\ge2$ and $|k|\ge2$, the effects of circulation due to vortex and chiral pairing can coexist, and hence depress the OAM simultaneously.  These observations can be explained by spectral asymmetry and unpaired fermions in the ground state of the BdG Hamiltonian. We also investigate the spatial distribution of particle density, OAM, by solving the BdG equation.

\end{abstract}

\maketitle

\section{Introduction}
\label{sec:Introduction}

In chiral superfluids (SFs), each Cooper pair carries a quantized relative orbital angular momentum (OAM)~\cite{Anderson1961, Leggett1975-RMP, Volovik-book-exotic, Read-Green2000, Volovik-book, Leggett-book}, e.g., $L_z=\nu$, with simplified pairing symmetry $\sim (p_x+ip_y)^{\nu}$, where $\nu=1,2,3$ for chiral $p$, $d$, and $f$-waves. One interesting question is whether a chiral SF exhibits a nonzero macroscopic OAM, which was originally raised for the Anderson-Brinkman-Morel phase (A phase) of ${}^3\mathrm{He}$~\cite{ABM1973}. This is an important question and it has received intense investigation in the past decades~\cite{Volovik-book, Leggett-book, Ishikawa1977, McClure-PRL1979, Mermin-Muzikar,Kita-review,Volovik-JETP1995,Kita1998,Stone-PRB2004,Stone2008,Sauls-PRB,Machida-JPSJ2012,Tsutsumi-PRB2012}. However, different viewpoints and approximations give different estimations~\cite{Kita-review}. One observation is that, since each Cooper pair carries OAM $\nu$, the total OAM should be $L_z=\nu N/2$. Another viewpoint starting from the Fermi liquid is that, since only the fermions near the Fermi surface contribute to the superconducting state, $L_z$ should be suppressed as  $L_z=(\Delta_0/E_F)^{\gamma}\nu N/2$~\cite{Anderson1961,Cross1975}, where $\Delta_0$ is the pairing gap amplitude and $E_F$ is the Fermi energy. Furthermore, different approximations lead to varying orders of magnitude $\gamma$, constituting the so-called \textit{orbital angular momentum paradox} (see a review~\cite{Kita-review} and references therein). Recent studies within the Bogoliubov-de Gennes (BdG) framework show there is a fundamental difference between $p+ip$-wave chiral SF and non-$p$-wave chiral SFs, manifested most clearly in the OAM and edge current~\cite{Nie2015PRL,Huang2014PRB,Huang2015PRB,Nie2020PRB}. It is found that for $p+ip$ SF in the BCS regime, the total OAM $L_z=\nu N/2$ is the same as that in the BEC regime. Namely, each Cooper pair carries relative angular momentum $1$ and contributes additionally to the total OAM. While for non-$p$-wave SFs, such as for $d+id$ SF, in a sharp confining potential, $L_z=\nu N/2 \times O(\Delta_0/E_F)\ll N$, which is much reduced from its ``full" value in the BEC regime. Such a difference can be explained by spectral asymmetry (or spectral flow)~\cite{Volovik-book,Paranjape,Niemi,Stone198789}, unpaired fermions near the boundary carrying net OAM, which cancels the contributions from the paired fermions, and the ground-state wave function of the BdG Hamiltonian~\cite{strong-Bogoliubov}. These results are supported by several subsequent studies~\cite{Volovik-JETP2015,Ojanen2016PRB,Suzuki2016PRB}.

Such counterintuitive phenomena are also found in $s$-wave SFs carrying multiply quantized vortices (MQV)~\cite{Radzihovsky2017PRL}. When fast rotating Fermi gas trapped in an anharmonic potential, quantized vortices appear~\cite{Pelster2009}, which are important topological defects, e.g., in understanding the Berezinskii-Kosterlitz-Thouless transition~\cite{Berezinskii,KT-transition}. The vortices can be divided into two categories, singly quantized vortices (elementary unit vortices with vorticity $|k|=1$), and multiply quantized vortices (MQV) with vorticity $|k|>1$.
A recent study shows a similarly fundamental difference between singly quantized vortex and MQV in $s$-wave superfluids, which is also remarkably manifested in OAM~\cite{Radzihovsky2017PRL}. For $s$-wave weakly-paired SF carrying a singly quantized vortex ($|k|=1$) in the BCS regime, the total OAM is given by $L_z=kN/2$, which is the same as that in the BEC regime. Namely, the simple picture in the BEC limit, where the elementary vortex induces a quantized OAM of $k=\pm 1$ per molecule, also holds in the BCS limit. However, for $s$-wave SF carrying an MQV ($|k|=1$) in the BCS regime, the total OAM is significantly reduced from its full value $kN/2$ in the BEC regime, by an amount $\sim (k_F \xi)^2$, where $k_F$ is the Fermi momentum, and $\xi$ is the coherence length. The reduction is also a result of unpaired fermions, which carry OAM opposite to that carried by Cooper pairs~\cite{Radzihovsky2017PRL}.  However, different from vortex-free chiral SFs, the reductions depend on the vortex core structure, but are independent of any boundary effect~\cite{Radzihovsky2017PRL}.

Although MQV is not energetically favorable compared with separated single-vorticity vortices in a homogeneous bulk system, it is expected to be energetically stable in chiral $p$-wave superconductors~\cite{Garaud2015,Sauls_2009}. Double-quantum vortex was proposed and observed in superfluid ${}^3\mathrm{He}$-A phase~\cite{Volovik-JETP1977,Lounasmaa1999PNAS,Blaauwgeers2000}. Therefore, chiral superfluid plays an ideal platform to investigate these features with MQV~\cite{Mizushima2008PRL,Mizushima2010PRA}. Since $s$-wave superfluid with MQV shares a similar BCS ground state as that of vortex-free chiral SFs, it is interesting to consider chiral $(p+ip)^{\nu}$-wave superfluid with a single MQV, and investigate if the effects caused by vortices coexist or are compatible with that due to chiral pairing, and if there is any spectral asymmetry~\cite{Volovik-book,Paranjape,Niemi,Stone198789} brought by in-gap edge modes.
In this work, we explore the chiral SFs (e.g. $p+ip$-wave with $\nu=1$ and $d+id$-wave with $\nu=2$) carrying a quantized vortex including singly quantized vortex ($|k|=1$) and MQV ($|k|>1$), and focus on the discussion of OAM, spectral flow, real space distribution of fermion density and OAM distribution (or edge current equivalently).

The rest of this paper is organized as follows. In Sec.~\ref{sec:BdG}, we give a review of the ideal settings of the two-dimensional chiral SFs confined by axisymmetric potentials, including the definition of symmetrized pairing operator, and a summary of BdG theory. In Sec.~\ref{sec:Vortex}, we present the details in solving the BdG equation of chiral SFs with an axisymmetric MQV, especially the details of the generalized Bogoliubov transformation. We also present the definitions and derivations of some observables we are concerned with. For simplicity, we assume the vortex is located at the center of the disk. The numerical results of energy spectra, spatial distributions, and discussions are demonstrated in Sec.~\ref{sec:Results}. The paper is briefly summarized in Sec.~\ref{sec:Conclusion}.

\section{chiral fermionic superfluids}
\label{sec:BdG}

In this section, we give a brief review of the ideal settings of the two-dimensional (2D) chiral superfluids (SFs) at zero temperature, in the framework of Bogoliubov-de Gennes (BdG) theory.  We start with the Bardeen-Cooper-Schrieffer (BCS) mean-field Hamiltonian for chiral fermionic SF,
\be
\hat{H}_{\mathrm{BCS}}=\int d\vbr\Big[\sum_{\sigma=\uparrow,\downarrow}\psi_{\sigma}^{\dag}(\vbr)\hat{T}\psi_{\sigma}(\vbr)
+\psi^{\dag}_{\uparrow}(\vbr)\hat{\Delta}\psi^{\dag}_{\downarrow}(\vbr)+\mbox{H.c.}\Big],
\label{eq-BCS}
\ee
where $\psi_{\sigma}$($\psi_{\sigma}^{\dagger}$) is the fermionic annihilation(creation) operator of spin $\sigma=\uparrow,\downarrow$.
For simplicity, the orbital quantization axis is assumed to be normal to the surface of the two-dimensional superfluid, $\vd=(0,0,d_z)$.
Then a symmetrized pairing operator $\hat{\Delta}$ is given by~\cite{Ivanov,Sauls-PRB}
\be
\hat{\Delta}=\frac12\Big[(\hat{p}_x+i \hat{p}_y)^\nu\Delta(\vbr)+\Delta(\vbr)(\hat{p}_x+i \hat{p}_y)^\nu\Big],
\label{eq-del-symm}
\ee
with momentum operators $\hat{p}_x=-i\p_x$, $\hat{p}_y=-i\p_y$. Since $\vbr$ and ${\bf p}$ do not commute with each other, we assume a symmetrized order for the pairing operator $\hat{\Delta}$ as above. Recall that $\nu=1$ represents the $p+ip$-wave pairing, and $\nu=2$ corresponds to the $d+id$-wave pairing. For a uniform superfluid with rotation symmetry, $\Delta(r)$ can be taken as a constant when neglecting the oscillation near the boundary. And it can be taken as $\Delta(r)=\Delta_0 \tanh(r/\xi)$, when considering the influence of a vortex at the center, where $\xi$ is the coherence length.

The kinematic energy operator is given by
\be
\hat{T}=-\frac{\nabla^2}{2m_0}+V(\vbr)-\mu,
\ee
where $V(\vbr)$ is some possible confining potential, and $\mu$ is the chemical potential. In the presence of rotation symmetry, when $V(r)$ describes the infinite circular well or a disk with a specular wall, the eigenfunction of $\hat{T}$ is the Bessel function. If $V(r)$ describes the harmonic trap, the eigenfunction of $\hat{T}$ is the Laguerre Polynomial.

In this paper, we always assume that the populations of the two spin components are equal. The pairing potential is defined as
\be
\Delta(\vbr)=g\ep{\psi_{\downarrow}(\vbr)(\hat{p}_x-i\hat{p}_y)^{\nu}\psi_{\uparrow}(\vbr)},
\label{eq-pair-potential}
\ee
where $g$ is the bare coupling constant.

The BCS mean-field Hamiltonian in Eq.~\eqref{eq-BCS} can be diagonalized by the Bogoliubov transformation,
\be
\psi_{\uparrow}(\vbr)=\sum_n\Big[u_n(\vbr)\gamma_{n,1}-v^*_n(\vbr)\gamma^\dag_{n,2}\Big],\label{eq-Bogo}\\
\psi^{\dag}_{\downarrow}(\vbr)=\sum_n\Big[v_n(\vbr)\gamma_{n,1}+u^*_n(\vbr)\gamma^\dag_{n,2}\Big],
\label{eq-Bogo1}
\ee
in which $\gamma_{n,a}, \gamma_{n,a}^{\dagger}, (a=1,2)$ are quasi-particle operators introduced by Bogoliubov transformation.
The coefficients of the above transformation satisfy the orthonormal conditions
\be
\int d\vbr\Big[u^*_m(\vbr)u_n(\vbr)+v^*_m(\vbr)v_n(\vbr)\Big]=\delta_{mn}.
\label{eq-uv-orth}
\ee
In terms of $\gamma$ operator, the BCS Hamiltonian in Eq.~\eqref{eq-BCS} can be diagonalized as
\be
\hat{H}_{\mathrm{BCS}}=E_0+\sum_{n,\sigma}E_n\gamma^{\dag}_{n,\sigma}\gamma_{n,\sigma},
\ee
where $E_0$ is the ground-state energy. With the above diagonalized Hamiltonian, one has the following commutation relations:
\be
[\hat{H}_{\mathrm{BCS}},\gamma_{n,\sigma}]=-E_n\gamma_{n,\sigma},\quad
[\hat{H}_{\mathrm{BCS}},\gamma^{\dag}_{n,\sigma}]=E_n\gamma^{\dag}_{n,\sigma}.
\ee
Substituting Eq.~(\ref{eq-BCS}) and the inverse of Eq.~(\ref{eq-Bogo}) and (\ref{eq-Bogo1}) into the above commutation relations, and equating both sides, we arrive at the Bogoliubov-de Gennes equation of 2D chiral superfluid explicitly,
\be
\left(
  \begin{array}{cc}
    \hat{T} & \hat{\Delta} \\
    \hat{\Delta}^\dag & -\hat{T}
  \end{array}
\right)\left(
  \begin{array}{c}
    u_n(\vbr) \\
    v_n(\vbr)
  \end{array}
\right)=E_n\left(
  \begin{array}{c}
    u_n(\vbr) \\
    v_n(\vbr)
  \end{array}
\right).
\ee

The pair potential in Eq.~\eqref{eq-pair-potential} is in turn determined by $u_n(\vbr)$ and $v_n(\vbr)$, the eigenfunctions  of BdG equation as,
\be
&&\Delta(\vbr)=g\sum_{E_n<0} v^*_n(\vbr)(\hat{p}_x-i\hat{p}_y)^{\nu}u_n(\vbr)\nonumber\\
&&\quad -g\sum_{E_n>0} u_n(\vbr)(\hat{p}_x-i\hat{p}_y)^{\nu}v^*_n(\vbr).
\label{eq-del-self}
\ee
Since we mainly focus on the topological properties, such as edge modes, spectral asymmetry (or spectral flow), and some other qualitative features, we will make almost no use of Eq.~\eqref{eq-del-self} in the rest of this paper. For this reason, we will not search for a self-consistent solution of the BdG equation in the current paper.

\section{Vortex solutions}
\label{sec:Vortex}

In this section, we further simplify the ideal setting of 2D chiral superfluid as one confined by a specular wall, considering a solution of the $p+ip$-wave or $d+id$-wave BdG equation with a multiply quantized vortex (MQV) located at the center of the disk, where the rotation axis is assumed to be perpendicular to the disk. Such confining potential is given by the infinite circular well,
\be
&&V(\vbr)=\left\{
          \begin{array}{ll}
            0, & (|\vbr|<R) \\
            \infty, & (|\vbr|>R)
          \end{array}.
        \right.
\label{eq-potential}
\ee
It is convenient to use the polar coordinates  $\vbr=(r, \theta)$ in such system, where the Laplacian operator becomes,
\be
\nabla^2=\frac{1}{r}\frac{\p}{\p r}\Big(r\frac{\p}{\p r}\Big)+\frac{1}{r^2}\frac{\p^2}{\p^2\theta},
\ee
and area element $d\vbr=rdr d\theta$. The $p+ip$-wave and $d+id$-wave pairing operators can be expressed in polar coordinates as
\be
(\hat{p}_x\pm i\hat{p}_y)^\nu=-ie^{\pm i\nu\theta}\Big(\frac{\p}{\p r}\pm\frac{i}{r}\frac{\p}{\p\theta}\Big)^\nu,
\label{eq-pw}
\ee
where $\nu=1$ for $p+ip$-wave pairing and $\nu=2$ for $d+id$-wave pairing.

Considering a vortex with an integer vorticity $k$, which is located at the center of the disk, the gap function can be assumed in the following form,
\be
\Delta(\vbr)=\Delta(r)e^{i k\theta},
\ee
where $\Delta(r)=\Delta_0 \tanh(r/\xi)$, and $\xi$ is the coherence length. With the functional form of the gap function, we can expand $u_n(\vbr)$ and $v_n(\vbr)$ by the Bessel functions as follows,
\begin{align}
&u_n(\vbr)=\sum_l u_n^{(l)}(\vbr)=\sum_{l,m}U^{(l)}_{mn}\phi_{l+k+\nu,m}(\vbr),\label{eq-uv}\\
&v_n(\vbr)=\sum_l v_n^{(l)}(\vbr)=\sum_{l,m}V^{(l)}_{mn}\phi_{l,m}(\vbr),\label{eq-uv1}
\end{align}
where the expansion functions are given by
\be
\phi_{l,m}(\vbr)=N_{l,m}J_l\Big(z_{l,m}\frac rR\Big)\frac{1}{\sqrt{2\pi}}e^{il\theta},\nonumber
\ee
which is the wave function of a single electron in an infinite circular well.
Here $J_l(x)$ is the first kind $j$th order Bessel function, $z_{lm}$ is the $m$th zero point of the function $J_l(x)$, and $N_{l,m}=\sqrt{2}/\big(R J_{l+1}(z_{l,m})\big)$ is the normalization factor.
The expansion functions satisfy the boundary condition $\phi_{l,m}(R)=0$, which is required by the confining potential in Eq.~\eqref{eq-potential}.
They also satisfy the orthogonal relations as follows
\be
\int d\vbr\,\phi^*_{l,m}(\vbr)\phi_{l',m'}(\vbr)=\delta_{ll'}\delta_{mm'}.
\label{eq-phi-orth}
\ee

Then the BdG equation becomes a matrix eigenvalue equation. Since $l$ is a good quantum number in such ideal settings with rotation symmetry, the matrix can be split into diagonal blocks for different values of $l$. For a given $l$, the BdG equation can be written as
\be
\sum_{m'}\left[
  \begin{array}{cc}
    T_{l+k+\nu} & D_l \\
    D_l^\dag & -T_{l}
  \end{array}
\right]_{m m'}\left[
  \begin{array}{c}
    U^{(l)}_{m'n} \\
    V^{(l)}_{m'n}
  \end{array}
\right]=E^{(l)}_n\left[
  \begin{array}{c}
    U^{(l)}_{mn} \\
    V^{(l)}_{mn}
  \end{array}
\right].
\label{eq-BdG}
\ee
Since the Bessel functions are the eigenfunctions of $\nabla^2$, we find that the kinematic term is simply a diagonal matrix given by
\be
(T_l)_{m m'}=\Bigg[\frac{(z_{l,m}/R)^2}{2m_0}-\mu\Bigg]\delta_{m m'},
\ee
where $m,m'=1,2,\cdots,M$. The matrix elements of the pairing potential are given by
\be
(D_l)_{m m'}=\int_0^R rdr\,\phi^*_{l+k+\nu,m}(\vbr)\hat{\Delta}\phi_{l,m'}(\vbr).
\ee
Making use of Eq.(\ref{eq-pw}), we find that
\be
&&(\hat{p}_x\pm i\hat{p}_y)^{\nu}J_l\Big(z_{l,m}\frac rR\Big)e^{il\theta}\nonumber\\
=&&(\pm i)^{\nu}(\frac{z_{lm}}{R})^{\nu}J_{l\pm\nu}\Big(z_{lm}\frac rR\Big)e^{i(l\pm\nu)\theta}.
\ee
Here we have used the recursion relations of the Bessel functions as follows
\be
&&J'_n(x)-\frac{n}{x}J_n(x)=-J_{n+1}(x),\\
&&J'_n(x)+\frac{n}{x}J_n(x)=J_{n-1}(x).
\ee
After the $\theta$ integral has been completed, the $D_l$ matrix elements can be obtained by using the symmetrized pairing operator $\hat{\Delta}$ in Eq.~\eqref{eq-del-symm} as,
\begin{widetext}
\be
&&(D_l)_{m m'}=\frac{i^\nu}2 N_{l+k+\nu,m}N_{l,m'}
\Bigg[\Big(\frac{z_{lm'}}{R}\Big)^\nu\int_0^R rdr\Delta(r)J_{l+k+\nu}\Big(z_{l+k+\nu,m}\frac rR\Big)J_{l+\nu}\Big(z_{l,m'}\frac rR\Big)\nonumber\\
&&\qquad+\Big(\frac{z_{l+k+\nu,m'}}{R}\Big)^\nu
\int_0^R rdr\Delta(r)J_{l+k}\Big(z_{l+k+\nu,m}\frac rR\Big)J_{l}\Big(z_{l,m'}\frac rR\Big)\Bigg].
\label{eq-D-matrix}
\ee
\end{widetext}
The integral in Eq.~\eqref{eq-D-matrix} can be calculated by Gaussian quadrature.
One can verify that the matrix $D_l$ satisfies the following relation,
\be
D_l=D_{-l-k-\nu}^T,
\ee
so that the particle-hole symmetry is valid in our model. More explicitly, if we denote the BdG matrix of a given $l$ as \be
H^{(l)}=
\left(
  \begin{array}{cc}
    T_{l+k+\nu} & D_l \\
    D_l^\dag & -T_{l}
  \end{array}
\right)_{m m'},
\ee
the particle-hole symmetry will map $H^{(l)}$ to $-H^{(-l-k-\nu)}$. Therefore, the energy spectrum is particle-hole symmetric about $l=-(k+\nu)/2$:
\be
\left\{E^{(l)}_{m} \right\}_{m\in {\mathbb N}} =\left\{ - E^{(-l-k-\nu)}_m \right\}_{m\in {\mathbb N}} .
\label{eq-PHS}
\ee

Since level crossing largely happens in such a system with in-gap edge modes, during which process the spectral flow might appear, it is more convenient to treat the particles and holes on the same footing. Namely, we do not distinguish the particle or hole operator in the Bogoliubov transformation. Instead, the meaning of creation or annihilation is determined by the ground state (see below Eq.~\eqref{eq-def-beta}).
To this end, we introduce the following generalized Bogoliubov transformation~\cite{strong-Bogoliubov} as
\be
&&\psi_{\uparrow}(\vbr)=\sum_{n,l} u^{(l)}_{n}(\vbr)\beta_{nl,1},\label{eq-gbogo1}\\
&&\psi^\dag_{\downarrow}(\vbr)=\sum_{n,l} v^{(l)}_{n}(\vbr)\beta_{nl,2},\label{eq-gbogo2}
\ee
where $u^{(l)}_{n}$ and $v^{(l)}_{n}$ are defined in Eqs.~\eqref{eq-uv} and ~\eqref{eq-uv1}, which are eigenfunctions for a given $l$.
Here $\beta_{nl,a} (a=1,2)$ are quasi-particle Bogoliubov operators. The ground state expectation is given by
\be
\ep{\beta^\dag_{nl,a}\beta_{nl,a}}=\left\{
                           \begin{array}{ll}
                             0, & E^{(l)}_n>0 \\
                             1, & E^{(l)}_n<0
                           \end{array},
                         \right.\quad\mbox{for}\quad a=1,2.
\label{eq-def-beta}
\ee
When $E^{(l)}_n>0$, $\beta_{nl,a}$ is an annihilation operator. While $E^{(l)}_n<0$, $\beta_{nl,a}$ is a creation operator.

By definitions of operators, generalized Bogoliubov transformation in Eq.~\eqref{eq-gbogo1},~\eqref{eq-gbogo2}, and orthonormal conditions of $u_n^{(l)}(\vbr)$ and $v_n^{(l)}(\vbr)$ in Eq.~\eqref{eq-uv-orth}, we can derive the expectation values of the following observables in the ground state. For example, the fermion density can be expressed as,
\be
&&n(\vbr)=\sum_\sigma\ep{\psi_\sigma^\dag(\vbr)\psi_\sigma(\vbr)}\nonumber\\
&&=\sum_l\Big[\sum_{E^{(l)}_n<0}|u^{(l)}_{n}(\vbr)|^2+\sum_{E^{(l)}_n>0}|v^{(l)}_{n}(\vbr)|^2\Big].
\label{eq-num}
\ee

Since a vortex or chiral superfluid is accompanied by circulating currents, we can also evaluate the azimuthal  mass current distribution,
\be
J_\theta(\vbr)=-i\frac1{m_0 r}\sum_{\sigma}\ep{\psi^{\dag}_{\sigma}(\vbr)\frac{\p}{\p \theta}\psi_{\sigma}(\vbr)}.
\ee
Similarly, we can also define the orbital angular momentum distribution as
\be
L_z(\vbr)=-i\sum_{\sigma}\ep{\psi^{\dag}_{\sigma}(\vbr)\frac{\p}{\p \theta}\psi_{\sigma}(\vbr)}.
\ee
Because these two quantities are related by $L_z(\vbr)=m_0 r J_\theta(\vbr)$, in the following discussions, we will focus on the angular momentum distribution $L_z(\vbr)$, which is equivalent to the circulating currents distribution $J_\theta(\vbr)$.

In the same way, the ground state expectation value of orbital angular momentum distribution can be expressed as,
\be
&&L_z(\vbr)=\sum_{l} \Big[(l+\mu+\nu)\sum_{E^{(l)}_n<0}|u^{(l)}_{n}(\vbr)|^2\nonumber\\
&&\qquad-l\sum_{E^{(l)}_n>0}|v^{(l)}_{n}(\vbr)|^2\Big].
\ee
We can integrate the local densities to obtain the total particle number and total orbital angular momentum as
\be
N=\ep{\hat{N}}=\int d\vbr\, n(\vbr),\quad L_z=\ep{\hat{L_z}}=\int d\vbr\, L_z(\vbr),
\ee
where OAM operator is given by $\hat{L}_z=-i\int d\vbr\sum_{\sigma}\psi^{\dag}_{\sigma}(\vbr)\p_{\theta}\psi_{\sigma}(\vbr)$, and total fermion number operator corresponds to $\hat{N}=\int d\vbr\sum_{\sigma}\psi^{\dag}_{\sigma}(\vbr)\psi_{\sigma}(\vbr)$.

Due to the pairing terms and the presence of vortices in the BCS Hamiltonian in Eq~\eqref{eq-BCS}, neither $\hat{L}_z$ nor $\hat{N}$ commutes with $\hat{H}_{\mathrm{BCS}}$. Therefore, neither one is conserved separately. However, one can define a generalized orbital angular momentum,
\be
\hat{Q} \equiv \hat{L}_z-\frac{k+\nu}{2}\hat{N},
\label{eq-hatQ}
\ee
and prove that $\hat{Q}$ commutes with $\hat{H}_{\mathrm{BCS}}$, which implies the expectation value $Q=\ep{\hat{Q}}$ is a conserved quantity~\cite{Volovik-JETP1995,Volovik-book,Radzihovsky2017PRL}. Physically, the generalized OAM operator evaluates the deviation of OAM from its ``full" value $L_z=(k+\nu)N/2$. So we can evaluate $Q$ by the eigenvalues of $\hat{Q}$ in the ground state of the BdG Hamiltonian, which are not trivial quantities.

By means of generalized Bogoliubov transformation in Eq.~\eqref{eq-gbogo1} and ~\eqref{eq-gbogo2}, we have proved in Appendix~\ref{app-A} that the expectation value $Q$ can be expressed by use of spectral asymmetry or spectral flow~\cite{Volovik-book,Paranjape,Niemi,Stone198789} $\eta_l$ as,
\be
Q=-\sum_l \Big(l+\frac{k+\nu}{2}\Big)\frac{\eta_l}2,
\label{eq-Q}
\ee
where $\eta_l$ is defined as
\be
\eta_l=\sum_{E^{(l)}_{n}>0}1-\sum_{E^{(l)}_{n}<0}1 =\sum_n \mbox{sgn} E_n^{(l)}.
\ee
If one finds $\eta_l=0$ for any $l$, then the total OAM takes the ``full" value as $L_z=(k+\nu)N/2$. Otherwise, any nonzero spectral flow $\eta_l$ introduces the suppression of OAM in the ground state.

Finally, we estimate the Fermi momentum and Fermi energy involved in this work. Integrating Eq.~(\ref{eq-num}) over the whole disk gives the total particle number $N$ of the Fermi superfluid. Assuming that the same amount of free fermions is contained in the same disk, then we find the following relation:
\be
\frac{N}{\pi R^2}=\frac{2\pi k_F^2}{(2\pi)^2}
\ee
Therefore, the Fermi momentum on a disk is given by $k_F^2=2N/R^2$, and the corresponding Fermi energy is $E_F=k^2_F/(2m_0)$. In the numerical calculations, we take $2m_0=1$ for convenience.

\section{Result and discussion}
\label{sec:Results}

In this section, we discuss the numerical solutions of BdG equations with a vortex inside the chiral superfluid. In solving the BdG equations in Eq.~\eqref{eq-BdG}, we assume the radius dependence of the pairing potential is given by
\be
\Delta(r)=\Delta_0\tanh\Big(\frac{r}{\xi}\Big),
\label{eq-del}
\ee
where $\Delta_0$ represents the overall scale of the gap amplitude, and $\xi=k_F/\Delta_0$ is the coherence length. We do not seek a self-consistent solution in this paper at present, since the qualitative features of the solutions do not depend on the detailed function form of $\Delta(r)$. In this sense, it is good enough to stick to the assumptions like Eq.~\eqref{eq-del} to discuss the qualitative behaviors of vortices in chiral superfluids.

\begin{figure}
\centering
\includegraphics[width=0.8\columnwidth]{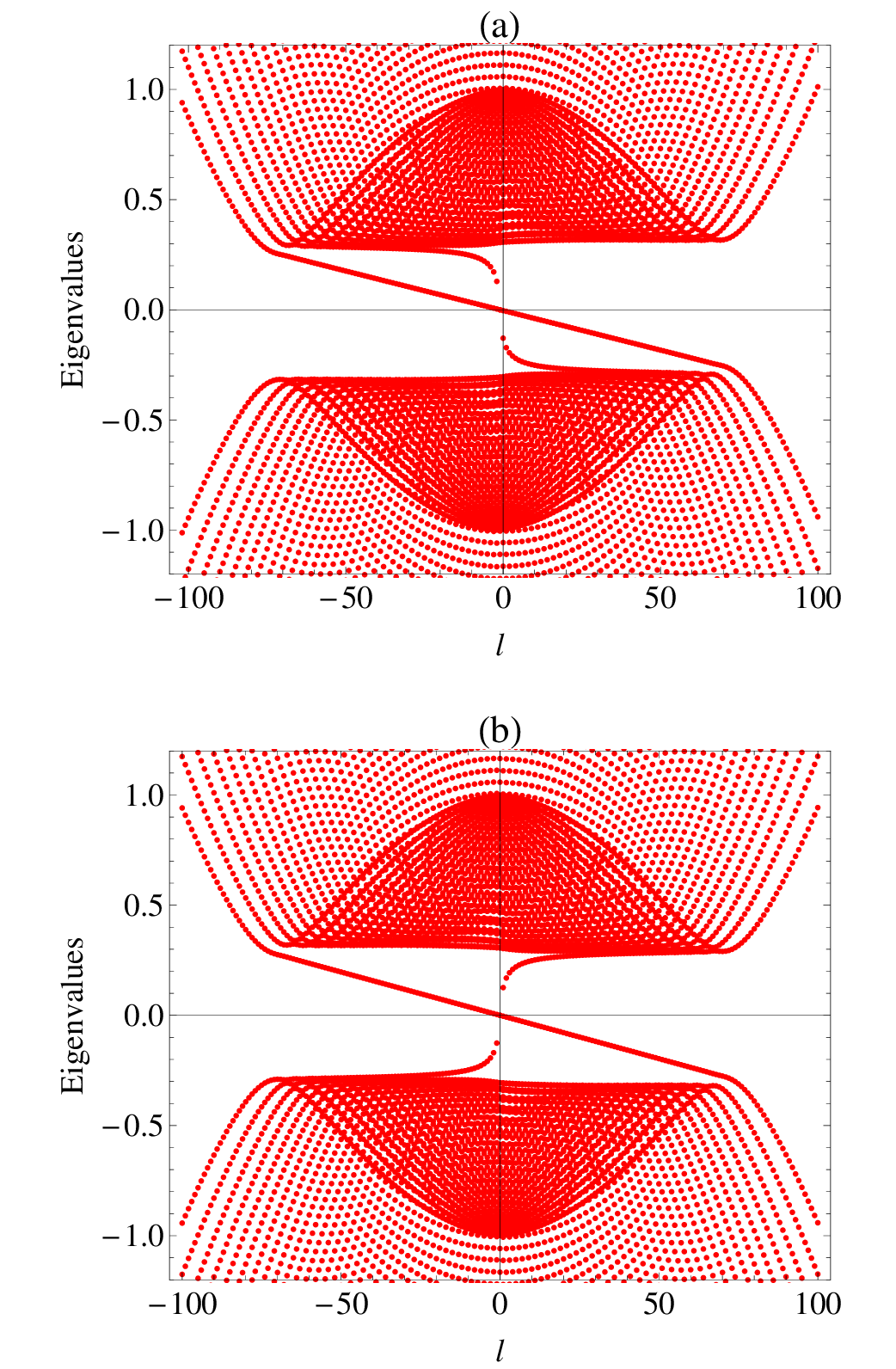}
\caption{The eigenvalues $E^{(l)}_n$ as a function of $l$. One singly quantized vortex is located at the center of the $p+ip$ chiral superfluid disk, with vorticity $k=1$ in (a) and $k=-1$ in (b), where $k_F R=80$, $k_F\xi=15$, $2m_0=1$, $\Delta_0/E_F=0.3$, $\mu/E_F=1$.}
\label{ev-v1p}
\end{figure}

In the BEC regime where the chemical potential $\mu/E_F<0$, which is a topologically trivial phase~\cite{Read-Green2000}, we find the energy spectrum is fully gapped without any in-gap state (not shown), for any integer vorticity $k$ and chiral pairing order $\nu$. In this case, the spectral flow is zero. According to Eq.~\eqref{eq-Q}, the generalized orbital angular momentum in the ground state $Q=0$ is found to hold for any $k$ and $\nu$. In other words, according to Eq.~\eqref{eq-hatQ}, the OAM of chiral SFs with a quantized vortex is given by $L_z=(k+\nu)N/2$ in the BEC regime, for any integer $k$ and $\nu$. Namely, each molecule contributes the OAM $l=k+v$, such that the total OAM is given by $L_z=(k+v)N/2$. However, in the BCS regime with $\mu/E_F>0$, which is a topological phase with in-gap states~\cite{Read-Green2000}, the situation could be different when level crossing happens. In the following, we focus on the discussion of energy spectrum and spatial distributions in the BCS regime.

In Figure~\ref{ev-v1p}, we compare the energy spectra of chiral $p+ip$-wave SFs with one singly quantized vortex with vorticity $k=1$ in (a), and that with an antivortex with vorticity $k=-1$ in (b). For simplicity, the vortex/antivortex is assumed to be located at the origin of the disk of chiral $p+ip$-wave SF. There are two chiral in-gap modes in both cases. The in-gap mode, which crosses a wider region with negative slope, is the edge mode due to $p+ip$ pairing, which is located at the disk boundary. The other in-gap mode that is more confined around $l=0$, corresponds to the edge mode located at the vortex core, which is so called Caroli–de Gennes–Matricon (CdGM) state~\cite{CdGM}. In Fig.~\ref{ev-v1p} (a), the slope of the vortex in-gap mode is also negative, since the circulation of the $k=1$ vortex is in the same direction as the $p+ip$ pairing. In Fig.~\ref{ev-v1p} (b), the vortex in-gap mode is reversed, because we have an antivortex here. In each case, the spectral flow $\eta_l$ is found to be zero for all $l$ (not shown). Therefore, according to Eq.~\eqref{eq-hatQ}, the generalized orbital angular momentum in the ground state $Q=0$ holds for each case. And according to Eq.~\eqref{eq-Q}, it implies $L_z=N$ for chiral $p+ip$ SFs with $k=1$ vortex, and $L_z=0$ for chiral $p+ip$ SFs with $k=-1$ antivortex, respectively.

\begin{figure}
\centering
\includegraphics[width=\columnwidth]{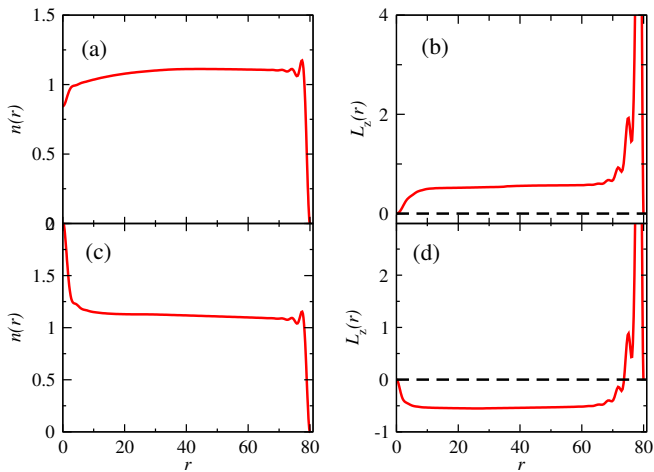}
\caption{The density $n(r)$ and the orbital angular momentum distribution $L_z(r)$ as functions of $r$, with one singly quantized vortex located at the the origin of chiral $p+ip$ -wave SF, where vorticity $k=1$ in the top row, and $k=-1$ in the bottom row.}
\label{dens-v1p}
\end{figure}

To have a more detailed understanding of the effects of vortices, we plot the density $n(r)$ and the orbital angular momentum (OAM) distribution $L_z(r)$ in Fig.~\ref{dens-v1p}. The results for chiral $p+ip$-wave SF with vorticity $k=1$ are presented in the top row. In Fig.~\ref{dens-v1p} (a), the density $n(r)$ has a small dip in the vortex core, which is caused by the circulation of the vortex. The OAM (or mass current) has a broad \textit{positive} plateau in the intermediate regime, which is also the contribution from the circulation of the vortex. Compared with the vortex-free $p+ip$ chiral SF, the plateau of $L_z(r)$ is around zero (not shown). And there is a large peak of the orbital angular momentum (OAM) near the boundary as shown in Fig.~\ref{dens-v1p} (b), which is caused by the edge modes due to $p+ip$ pairing. These two effects, coming from the vortex and $p+ip$ pairing, contribute together to give rise to the total OAM $L_z=N$. It coincides with the simple physical picture that all fermions are bound into $N/2$ Cooper pairs, and each of them contributes the OAM $l=k+\nu=2$. This explains that the total OAM is the overall contributions from all the Cooper pairs, such that we have $L_z=(N/2)l=N$.

In contrast, we also investigate the density $n(r)$ and OAM distribution $L_z(r)$ of a $k=-1$ antivortex in the $p+ip$ chiral SF in the lower panels in Fig.~\ref{dens-v1p}. In this case, the phase factor of $\Delta(\vbr)$ of the antivortex cancels that of $p+ip$ pairing. Although the overall OAM $L_z$ is exactly zero due to zero spectral flow, the distribution of OAM (or mass current) is non-trivial as shown in Fig.~\ref{dens-v1p} (d). The contribution from the broad \textit {negative} plateau due to the antivortex exactly cancels that from the positive peaks near the boundary due to $p+ip$ pairing. In the meantime, the density plot in Fig.~\ref{dens-v1p} (c) shows a small peak around the disk center, instead of a dip inside the vortex core, which is the contribution from the antivortex.
In summary, although the total OAM of $p+ip$ antivortex solution is the same as that of the uniform $s$-wave vortex-free SF, the OAM (or mass current) distribution is non-trivial.

\begin{figure}
\centering
\includegraphics[width=\columnwidth]{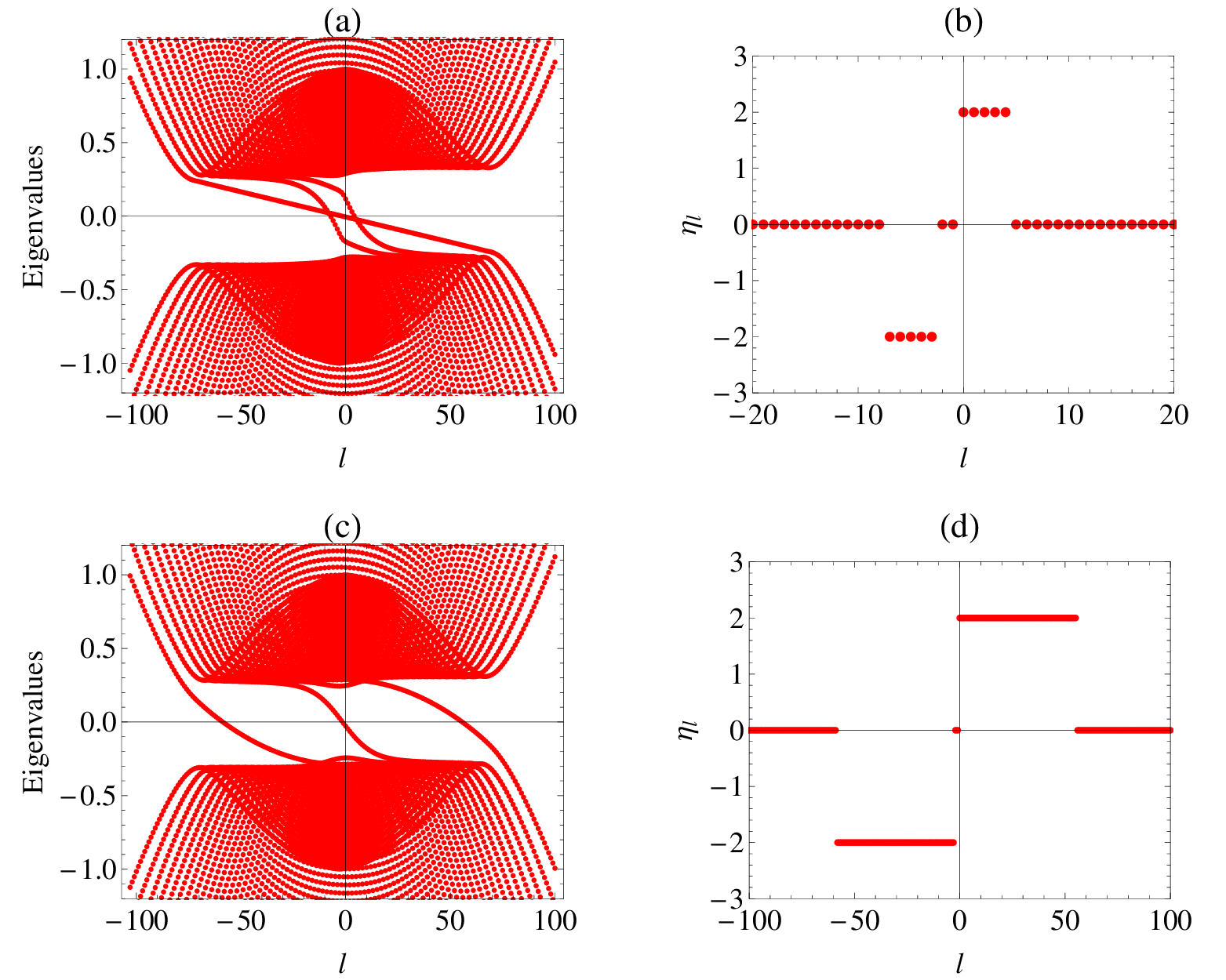}
\caption{The eigenvalues $E^{l}_n$ and spectral flow $\eta_l$ as functions of $l$. Top row: One $k=2$ vortex in the $p+ip$ chiral SF. Bottom row: One $k=1$ vortex in the $d+id$ chiral SF. Parameters: $k_F R=80$, $k_F\xi=15$, $2m_0=1$, $\Delta_0/E_F=0.3$, $\mu/E_F=1$.}
\label{ev-v2}
\end{figure}

Now we turn to the case of  $p+ip$ chiral SF with a multiply quantized vortex (MQV), where $k=2$. The energy spectrum $E^{(l)}_n$ and spectral flow $\eta_l$ as functions of $l$ are plotted in Fig.~\ref{ev-v2} (a) and (b), in which the particle-hole symmetry $E^{(l)}\sim-E^{(-l-k-\nu)}$ is preserved.
There are two narrow chiral in-gap dispersions located around $l\sim-(k+\nu)/2$, representing the edge modes located inside the vortex core. Other than these two narrow edge modes, a much broader chiral in-gap edge mode crosses the center of the spectrum, which is located near the system boundary due to the $p+ip$ pairing. We find the spectral flow $\eta_l$ in this case is nonzero for $0\leq l\leq 4$ and $-7\leq l\leq -3$ with given parameters, which completely comes from the contributions of the in-gap vortex modes, instead of from the edge mode due to $p+ip$ pairing. The region of the non-zero spectral flow caused by the vortex is quite narrow, compared with that caused by chiral pairing, as elaborated in the following.

In Fig.~\ref{ev-v2} (c) and (d), we plot the energy spectrum $E^{(l)}_n$ and spectral flow $\eta_l$ for $d+id$ chiral SF with a singly quantized vortex, where $k=1$. In this case, one in-gap vortex mode crosses the center of the energy spectrum, which is located at the center of the vortex core.  The other two wider chiral in-gap dispersions crossing around $l_{-}$ and $l_+$, are the edge modes due to $d+id$ pairing, where $|l_{\pm}|\sim k_F R$. In this case, the spectral flow is nonzero for $0\leq l\leq 55$ and $-58\leq l\leq -3$ for given parameters.  The nonzero spectral flow caused by the chiral $d+id$ pairing spans a much larger region, compared with that due to MQV (eg, in Fig~\ref{ev-v2} (b)).

\begin{figure}
\centering
\includegraphics[width=\columnwidth]{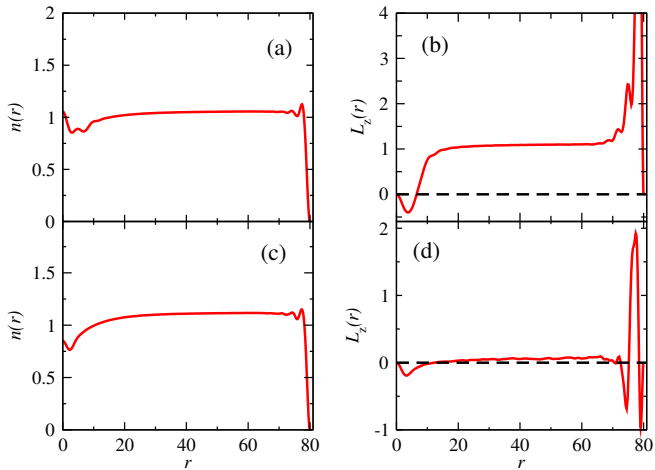}
\caption{The density $n(r)$ and the OAM distribution $L_z(r)$ as functions of $r$. Top row: One $k=2$ vortex in the $p+ip$ chiral SF. Bottom row: One $k=1$ vortex in the $d+id$ chiral SF. Parameters: $k_F R=80$, $k_F\xi=15$, $2m_0=1$, $\Delta_0/E_F=0.3$, $\mu/E_F=1$.}
\label{dens-v2}
\end{figure}

We further plot the density $n(r)$ and OAM $L_z(r)$ distributions for $p+ip$ SF with $k=2$ vortex, and $d+id$ SF with $k=1$ vortex in Fig.~\ref{dens-v2}. First we compare singly quantized vertex ($k=1$, Fig.~\ref{dens-v1p} (a-b)) with MQV ($k=2$, Fig.~\ref{dens-v2} (a-b)) in chiral $p+ip$ SF. We find that the $k=2$ vortex has a larger density dip than that with the $k=1$ vortex, indicating that higher vorticity generates a larger vortex core. Furthermore, counter flow appears in $p+ip$ SF with $k=2$ vortex, locating near the center of the vortex core, which is attributed to the unpaired fermions in the ground state. While for $p+ip$ SF with singly quantized vertex ($k=1$), the OAM (or current) distribution is positive definite in the whole region, manifesting the remarkable difference between singly quantized vertex and MQV in current distribution.
When comparing $p+ip$ SF and $d+id$ SF with a singly quantized vortex (Fig.~\ref{dens-v1p} (a-b) and Fig.~\ref{dens-v2} (c-d)), there is also a small negative peak appearing close to the boundary in the $d+id$ superfluid. The counterflow near the boundary is due to the two in-gap modes located at the system boundary, also attributed to the unpaired fermions in the ground state.

One can also see that there is a small negative peak appearing in the vortex core in Fig.~\ref{dens-v2} (d). This result is a little surprise since there is no spectral flow or unpaired fermions for a singly quantized vortex. We believe that the emergence of this small counterflow inside the vortex core is due to the interplay between the vortex and the high partial wave pairing. If we compare the density profile of Fig.~\ref{dens-v2} (c) and Fig.~\ref{dens-v1p} (a), one can see that the vortex core size is slightly larger in the $d+id$ pairing than in the $p+ip$ pairing. The larger core size means that $L_z(r)$ has a larger space to oscillate around zero and may generate some counterflows. But the existence of counterflow in the core of a $k=1$ vortex may not be robust and depends on the detailed shape of $\Delta(r)$.

Finally, let us discuss why the OAM is suppressed in the presence of spectral flow. A general expression of the ground state of a BdG Hamiltonian is given by $|{\rm GS}\rangle \sim \otimes_l |{\rm GS}\rangle_l$~\cite{strong-Bogoliubov},
\begin{align}
|{\rm GS}\rangle_l&=\left(\prod_{j=1}^{M_{\uparrow}^{(l)}}
\tilde{c}_{j,l+k+\nu,\uparrow}^{\dagger}\right)
\left(\prod_{j=1}^{M_{\downarrow}^{(l)}}
\tilde{c}_{j,-l,\downarrow}^{\dagger}\right)\notag\\
&\times
\exp \left(\sum_{j>M_{\uparrow}^{(l)}}^M\sum_{j^{\prime}>M_{\downarrow}^{(l)}}^M
\tilde{c}_{j,l+k+\nu,\uparrow}^{\dagger}
F^{(l)}_{jj^{\prime}}
\tilde{c}_{j^{\prime},-l,\downarrow}^{\dagger}\right)|0\rangle,
\label{eq:GS}
\end{align}
where $|0\rangle$ is the vacuum state, $\tilde{c}_{j,l,\sigma}$ is the linear superpositions of Fermion operator $c_{j,l,\sigma}$ which is define by the expansion of field operator as $\psi_\sigma(\vbr)=\sum_{j,l}c_{j,l,\sigma}\phi_{j,l}(\vbr)$. The number of un-paired fermion is computed by $M_{\uparrow,\downarrow}^{(l)}=\mbox{max}(0,M-M_{+,-}^{(l)})$, where $M_{+,-}^{(l)}$ is the number of positive (negative) eigenvalues of BdG matrix with given $l$, and $2M$ is the dimension of the reduced BdG matrix $H^{(l)}$. When $M_{+}^{(l)}=M_{-}^{(l)}=M$, or $M_{\uparrow,\downarrow}^{(l)}=0$, Eq.~\eqref{eq:GS} is reduced to the BCS ground state wavefunction, where all the fermions are paired into Cooper pairs. In general, $M_{+}^{(l)}$ is not necessarily equivalent to $M_{-}^{(l)}$, especially in the system with in-gap edge modes, where level crossings largely happen. And the difference between them yielding the spectral asymmetry $\eta_l=M_{+}^{(l)}-M_{-}^{(l)}$. Furthermore, it can be proved that $\eta_l=2(M_{\downarrow}^{(l)}-M_{\uparrow}^{(l)})$, where nonzero $M_{\uparrow,\downarrow}^{(l)}$ denotes the unpaired fermions in the energy levels after a generalized Bogoliubov transformation~\cite{strong-Bogoliubov,Radzihovsky2017PRL}. It implies that nonzero spectral flow signals the existence of unpaired fermions, which contribute to the reduction of the OAM~\cite{Nie2015PRL,Radzihovsky2017PRL}.

\section{Conclusion}
\label{sec:Conclusion}

In conclusion, we have presented a detailed study of a variety of vortices in the chiral superfluid with different partial-wave pairing. It is found that in the $k=\pm1$ vortices of the $p+ip$ chiral superfluid, there is no spectral asymmetry and the OAM exactly equals the contributions of all Cooper pairs. In the $k=-1$ case, the spatial OAM distribution is nontrivial although the total OAM is exactly zero. In contrast, the spectral flows emerge when the vorticity is higher than $1$ or the partial-wave pairing is higher than $p+ip$, which leads to the suppression of the total OAM. The effects of MQV or high partial-wave pairing on the spectral flows are almost independent of each other. In these cases, some counter-flow currents due to the unpaired fermions are also observed either in the vortex core or near the system boundary. In a qualitative sense, these counterflows explain the decrease of the OAM from the naive expected value.

\section{Acknowledgement}
W. N. was supported by the Natural Science Foundation of China under Grant No. 12174273. Y. H. was supported by the Natural Science Foundation of China under Grant No. 11874272.
\appendix
\begin{widetext}
\section{Some detail derivations}
\label{app-A}

In this appendix, we present the proof that
\be
Q=L_z-\frac{k+\nu}{2}N=-\sum_l \Big(l+\frac{k+\nu}{2}\Big)\frac{\eta_l}2.
\ee
Considering the expansion of $u^{(l)}_n(\vbr)$ and $v^{(l)}_n(\vbr)$ in Eq.~\eqref{eq-uv} and ~\eqref{eq-uv1}, we find that
\be
Q&=&\sum_l \Big(l+\frac{k+\nu}{2}\Big)\int d\vbr\Big[\sum_{E^{(l)}_n<0}|u^{(l)}_{n}(\vbr)|^2-\sum_{E^{(l)}_n>0}|v^{(l)}_{n}(\vbr)|^2\Big]\nonumber\\
&=&\sum_l \Big(l+\frac{k+\nu}{2}\Big)\Big[\sum_{m,E^{(l)}_n<0}\Big(U^{(l)}_{mn}\Big)^2-\sum_{m,E^{(l)}_n>0}\Big(V^{(l)}_{mn}\Big)^2\Big].
\ee
Here we have used the orthogonal relations of $\phi_{lm}(\vbr)$ in Eq.~\eqref{eq-phi-orth}.
Now it is easy to see that the following matrix
\be
\left(
  \begin{array}{c}
    U^{(l)}_{mn} \\
    V^{(l)}_{mn}
  \end{array}
\right),\quad\mbox{for}\quad m=1,\cdots,M,\quad n=1,\cdots,2M,
\ee
is an orthogonal matrix. Therefore, each row and column of the above matrix is a unit vector,
\be
\sum_{m=1}^{M}\Big[\Big(U^{(l)}_{mn}\Big)^2+\Big(V^{(l)}_{mn}\Big)^2\Big]=1,\quad
\sum_{n=1}^{2M}\Big(U^{(l)}_{mn}\Big)^2=1,\quad \sum_{n=1}^{2M}\Big(V^{(l)}_{mn}\Big)^2=1.
\ee
Making use of the above identities, we find that
\be
&&\sum_{m,E^{(l)}_n<0}\Big(U^{(l)}_{mn}\Big)^2-\sum_{m,E^{(l)}_n>0}\Big(V^{(l)}_{mn}\Big)^2
=\sum_{m,E^{(l)}_n<0}\Big(U^{(l)}_{mn}\Big)^2-\sum_{E^{(l)}_n>0}\Big[1-\sum_m\Big(U^{(l)}_{mn}\Big)^2\Big]\nonumber\\
&&=\sum_{m,n}\Big(U^{(l)}_{mn}\Big)^2-\sum_{E^{(l)}_n>0}1=M-\sum_{E^{(l)}_n>0}1=-\frac{1}{2}\eta_l.
\ee
Here we have defined the spectral flow,
\be
\eta_l=\sum_{E^{(l)}_{n}>0}1-\sum_{E^{(l)}_{n}<0}1.
\ee
Collecting the above results, we find that $Q$ depends on the spectral flows as
\be
Q=-\sum_l \Big(l+\frac{k+\nu}{2}\Big)\frac{\eta_l}2 =\sum_n \mbox{sgn} E_n^{(l)}.
\ee
If there is no spectral flow, then we have $L_z=(k+\nu)N/2$.

\end{widetext}


%

\end{document}